\documentclass[aps,prd,twocolumn,groupedaddress,amssymb,amsmath,nofootinbib]{revtex4}

\usepackage{color}
\usepackage{graphicx}
\usepackage{epstopdf}
\usepackage{hyperref} 

\begin{document}

\preprint{}

\title{
Loop suppressed electroweak symmetry breaking and naturally heavy superpartners 
}

\author{Radovan Derm\' \i\v sek}

\affiliation{Physics Department, Indiana University, Bloomington, IN 47405, USA}
\affiliation{Department of Physics and Astronomy
and Center for Theoretical Physics, Seoul National University, Seoul 151-747, Korea}

\email[]{dermisek@indiana.edu}


\date{November 28, 2016}

\begin{abstract}

A model is presented in which ${\cal O} (10 {\rm \; TeV})$ stop masses, typically required by the Higgs boson mass in supersymmetric models, do not originate from soft supersymmetry breaking terms that would drive the Higgs mass squared parameter to large negative values but rather from the mixing with vectorlike partners. Their contribution to  the Higgs mass squared parameter is reduced to  threshold corrections and, thus, it is one loop suppressed compared to usual scenarios. 
New fermion and scalar partners of the top quark with ${\cal O} (10 {\rm \; TeV})$ masses are predicted.

\end{abstract}


\maketitle






\section{Introduction}

Electroweak~symmetry~breaking (EWSB)  is very elegant in supersymmetric models. It is radiatively driven by the top Yukawa coupling and the electroweak (EW) scale is tightly related to masses of  superpartners of the top quark (stops) propagating in the loops. 

However, the most straightforward explanation of the measured value of the Higgs boson mass, $m_h \simeq 125$ GeV, suggests at least ${\cal O}(10 \;{\rm TeV})$ stop masses~\cite{Hahn:2013ria, Draper:2013oza} and, in such scenarios, generating a two orders of magnitude smaller EW scale  requires  tremendous fine-tuning, at least 1 part in $10^4$, in relevant parameters. 
It might be possible to avoid this little hierarchy problem if a model is built with specific relations between soft supersymmetry (SUSY) breaking parameters that lead to required cancellations 
 or that generate large additional contributions to the Higgs boson mass,  such as contributions from stop mixing in the minimal supersymmetric model (MSSM) or from new couplings in models beyond the MSSM. Nevertheless, avoiding large fine-tuning in EWSB  requires  significantly more  complex models  or stretching the parameters far beyond what was considered reasonable before the Higgs discovery (and often giving up some desirable features, like perturbativity to a high scale)~\cite{EWSB_reviews}.

In this paper, a solution  is presented in which  ${\cal O} (10 {\rm \; TeV})$ stop masses do not originate from soft SUSY breaking terms that would drive the Higgs mass squared parameter, $\tilde m_{H_u}^2$, to large negative values but rather from the mixing with vectorlike partners. Therefore, an arbitrarily small contribution to $\tilde m_{H_u}^2$ is generated from the Yukawa coupling to scalars in the renormalization group (RG) evolution from a high scale. The contribution from scalars   is reduced to  threshold corrections and higher-order effects. Thus, it is one loop suppressed compared to  usual scenarios allowing for more natural EWSB.
 
\begin{figure}[b]
\includegraphics[width=2.8in]{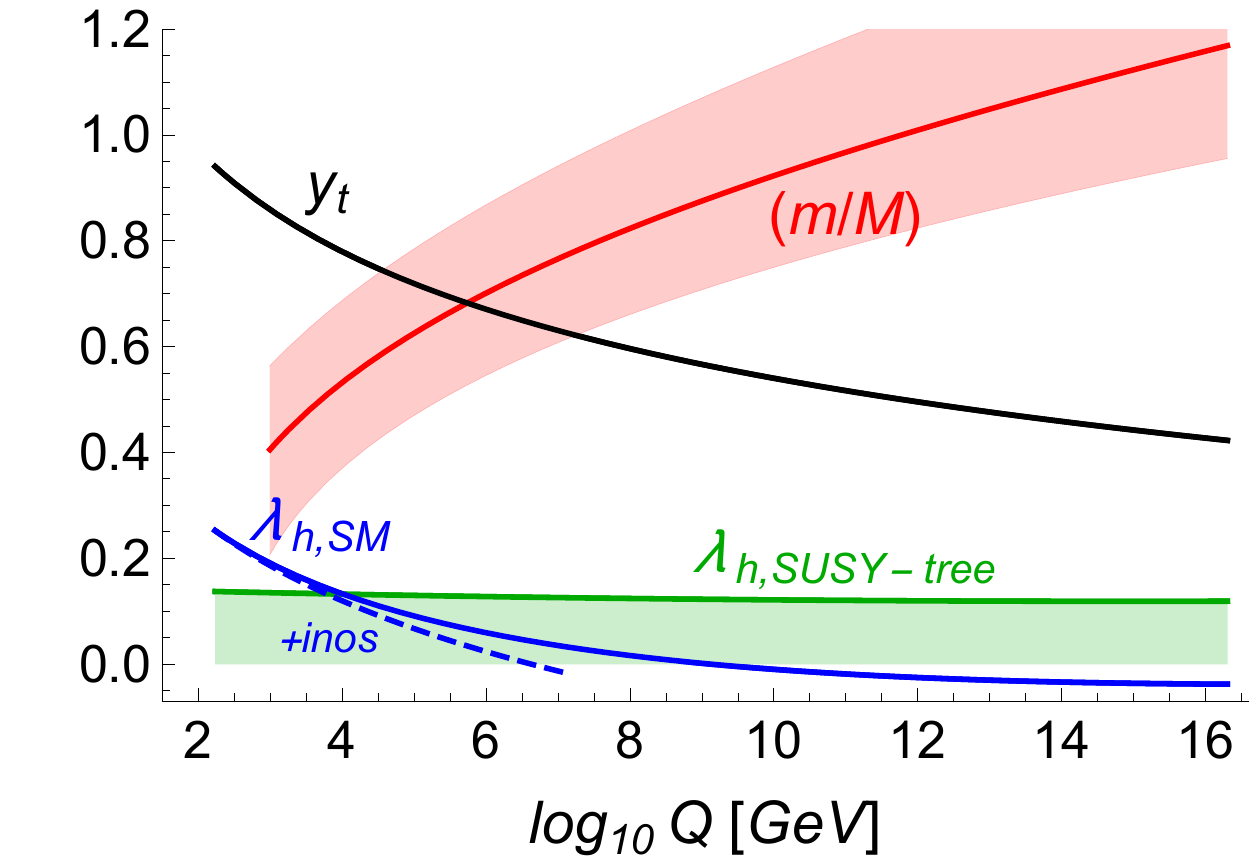}
\caption{RG evolution of top Yukawa coupling, $y_t$, the Higgs quartic coupling in the SM, $\lambda_{h, SM}$, and in the SM with electroweak gauginos and Higgsinos (indicated by dashed line). RG evolution of the tree level prediction for $\lambda_{h}$ in the MSSM is shown in shaded region with the solid line representing its maximum value, $\lambda_{h,SUSY-tree} = (g_2^2+g_Y^2)/4$. The  $m/M$ line and shaded region indicate the value required to obtain the correct $y_t(Q)$ for $\lambda = 1\pm 0.1$  at $Q= (M^2+m^2)^{1/2}$.
}
\label{fig:RG}
\end{figure}

 The need for heavy stops can be seen from the approximate analytic formula for the Higgs boson mass, 
       \begin{equation}
 m_{h}^2 \simeq M_Z^2 +
  \frac{3 y_t^2}{4\pi^2}  m_{t}^2   \ln \left( \frac{m_{\tilde t}^2}{ m_{t}^2} \right),
  \label{eq:mh}
  \end{equation}
assuming  medium or large $\tan \beta$ (the ratio of vacuum expectation values of the two Higgs doublets) in which case the tree level result (the first term) is maximized. Alternatively, it can be seen in the plot of the RG evolution of the Higgs quartic coupling in the standard model (SM) and its tree level prediction in the MSSM given by $SU(2)$ and $U(1)_Y$ gauge couplings, $\lambda_{h,SUSY-tree} = (g_2^2+g_Y^2)/4$. From Fig.~\ref{fig:RG}, we see that they intersect at about 10 TeV which indicates the scale at which  superpartners should be integrated out to obtain the measured value of the Higgs mass. 
The exact stop masses needed depend on the assumptions for  masses of gauginos and Higgsinos (collectively called  ``inos"), with light inos favoring smaller stop masses, as indicated by dashed line in Fig.~\ref{fig:RG}. We use two-loop RG equations summarized in Refs.~\cite{Binger:2004nn, Giudice:2011cg, Degrassi:2012ry, Buttazzo:2013uya}~\cite{Draper:2013oza}.

Large soft trilinear couplings, $A$ terms,  result in stop mixing which  modifies 
Eq.~(\ref{eq:mh}); analogous formula can be found in Ref.~\cite{Carena:1995bx}. These contributions  can  also be viewed as threshold corrections to Higgs quartic coupling that modify the tree level prediction and alter the scale at which SUSY should be matched to the SM in Fig.~\ref{fig:RG}. However, large threshold corrections require specific relations between parameters, far from typically obtained in SUSY models. In this paper, we focus on generic spectrum that typically leads to small threshold corrections.

 The mass of the $Z$ boson  in the MSSM, away from small $\tan \beta$ regime, is given by:
   \begin{equation}
M_Z^2 \simeq - 2\mu^2(M_Z) - 2\tilde m_{H_u}^2(M_Z) ,
\label{eq:mh-ewsb}
  \end{equation}
  where $\mu$ is the supersymmetric Higgs mass parameter.
 Heavy stops contribute to the RG running of  $\tilde m_{H_u}^2$:
  \begin{equation}
  \frac{d \;\tilde m_{H_u}^2}{d \ln Q} = 
 \frac{3 y_t^2}{8\pi^2} \left( \tilde m^2_{H_u} + \tilde m^2_{ t_L} + \tilde m^2_{ t_R} \right) ,
 \label{eq:rge}
  \end{equation}
 where we neglected  contributions from gaugino masses and $A$ terms. In this approximation, stop soft masses squared, $\tilde m^2_{ t_L}$ and $\tilde m^2_{t_R}$, have the same RG equations up to overall factors 1/3 and 2/3 respectively. 
 The typical outcome  of the  RG evolutions from a high scale is $\tilde m^2_{H_u} \simeq - ( \tilde m^2_{ t_L} + \tilde m^2_{ t_R})$
 and, for  ${\cal O} (10 {\rm \; TeV})$ stops, it results in already mentioned $\sim 0.01\%$  tuning required in Eq.~(\ref{eq:mh-ewsb}). More importantly, however, a large contribution is already generated  in 
 the RG evolution over one decade in the energy scale, requiring $\sim 0.1\%$ tuning.

Besides stop masses, a significant fine-tuning can also result from the gluino mass. Although gluino doesn't couple to $H_u$ directly, it drives stop masses to positive values which in turn drive $\tilde m_{H_u}^2$ to negative values. Solving coupled RG equations, we find that current limits on gluino mass, ${\cal O} (1 {\rm \; TeV})$, result in $\sim 1\%$ tuning in EWSB for high-scale mediation scenarios. Alternatively, not larger than   $\sim 10\%$ tuning allows for about three decades of  RG evolution and, thus, favors models with low-scale mediation of SUSY breaking. 

While limits on gluino do not necessarily prevent building a  model with natural EWSB without  specific relations between parameters, ${\cal O} (10 {\rm \; TeV})$ stops make it impossible in models like MSSM even for a low mediation scale.  In the model that follows, the $\tilde m_{H_u}^2$ does not run at one-loop level due to scalar masses irrespectively of the mediation scale.

\section{ Model}

Part of the superpotential related to the top quark is given by:
\begin{equation}
W \supset  \lambda  q \bar u H_u + m_{q} q \bar Q + m_{u} U \bar u + M_{Q} Q \bar Q + M_{U} U \bar U,
\label{eq:W}
\end{equation}
where $q$ and  $\bar u$, collectively called $f$, have the quantum numbers of SU(2) doublet and  singlet up-type  quarks in the MSSM. The $H_u$ is the Higgs doublet that couples to up-type quarks, and $\lambda$ would be the usual top Yukawa coupling if there was no mixing with vectorlike quarks. 
Capital letters denote extra vectorlike pairs  that do not couple directly to the $H_u$; $Q$ and  $\bar U$, collectively called $F$, ($\bar Q$ and  $U$,  collectively called $\bar F$) have the same (opposite) quantum numbers as 
$q$ and  $\bar u$.

Although the explicit mass terms in Eq.~(\ref{eq:W}) are the most general consistent with SM gauge symmetries, the Yukawa couplings are not. However, presence of other couplings, if they are sufficiently small, does not alter our discussion and thus we neglect them. Alternatively, 
the explicit mass terms may originate from vevs of SM singlets $S_m$ and $S_M$: $m_{q, u} = \lambda_{q, u} \langle S_m \rangle$ and $M_{Q, U} = \lambda_{Q, U} \langle S_M \rangle$.  This allows us to  distinguish $F$ from $f$ by a $U(1)$ charge and  uniquely fix the structure of the superpotential in Eq.~(\ref{eq:W}). For example: $Q_F = +1$,  $Q_{\bar F} = -1$, $Q_{S_m} = +1$ with other fields not being charged. The same charges can be  extended to whole families. We will see that assuming this origin of vectorlike mass terms also allows for a natural connection between vectorlike masses and soft SUSY breaking  masses of corresponding scalars.

The mass matrix for fermions with $\pm 2/3$ electric charge  in the basis 
\begin{equation}
\left( 
\begin{array}{ccc}
q & Q & U
\end{array}
\right)
M_F 
\left( 
\begin{array}{c}
\bar u \\
\bar Q \\
\bar U
\end{array}
\right)
\end{equation}
is given by:
\begin{equation}
M_F  = 
\left( 
\begin{array}{ccc}
\lambda v_u & m_q & 0 \\
0 & M_Q & 0 \\
m_u & 0 & M_U \\
\end{array}
\right),
\label{eq:MF}
\end{equation}
where we use the same labels for the $\pm 2/3$ charge components of doublets as for  whole doublets (this should not result in any confusion since we only  discuss the sector related to top quark). The $v_u = v \sin \beta$ is 
 the vev of $H_u$ in a normalization with $v \simeq 175$ GeV.

\begin{figure*}[th]
\includegraphics[width=2.55in]{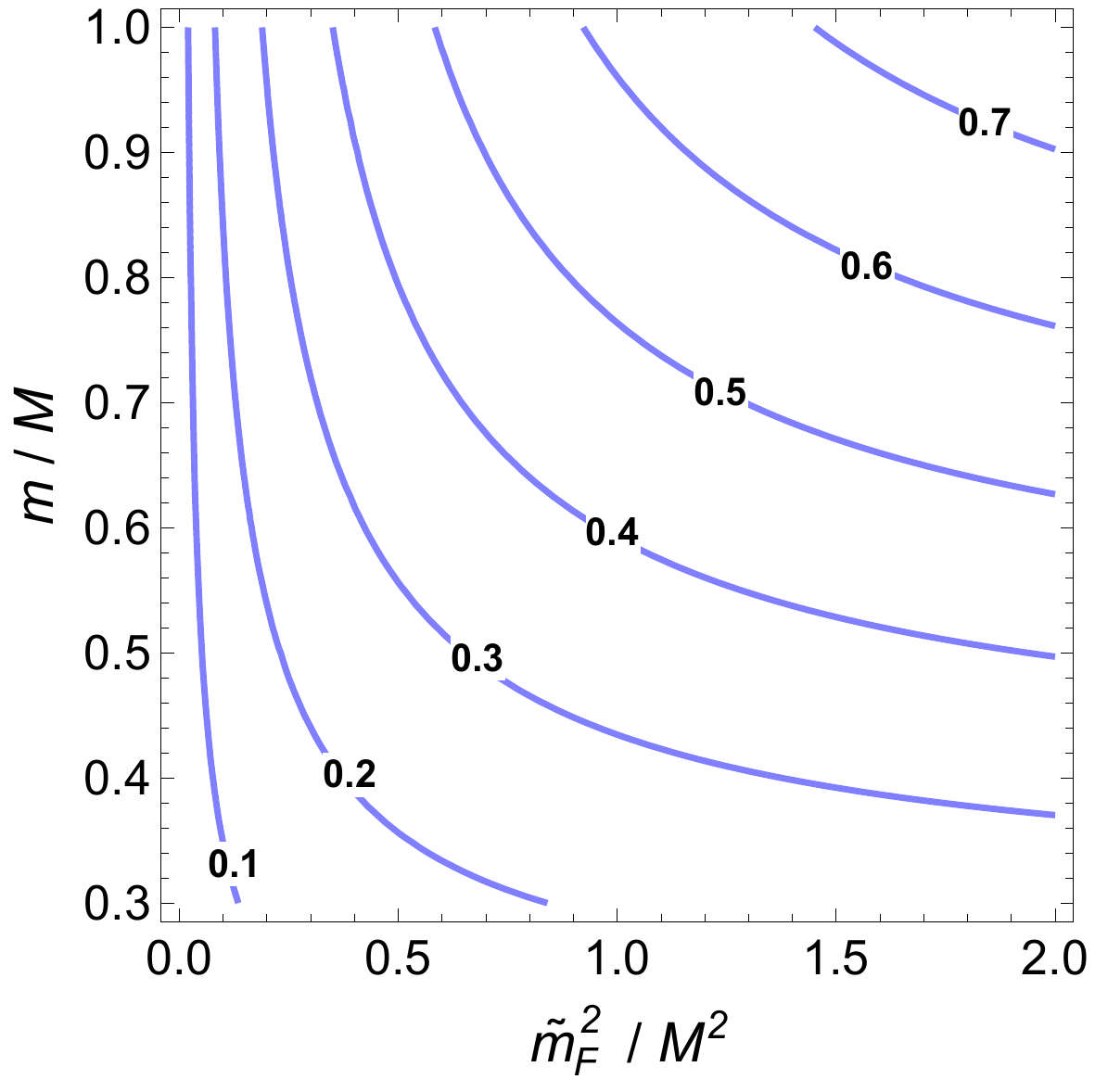}
\hspace{0.5cm}
\includegraphics[width=2.55in]{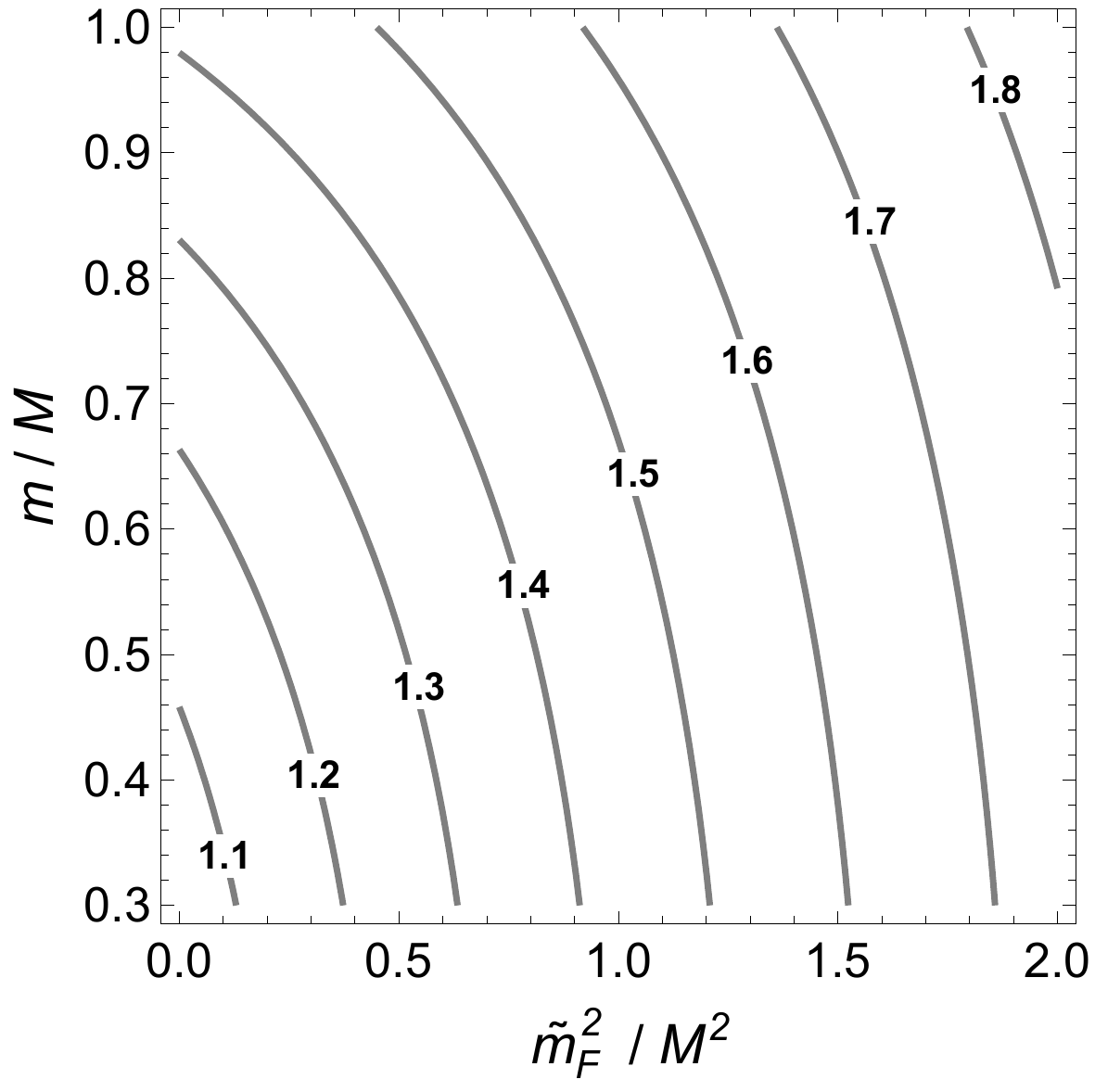}
\caption{The $m_{\tilde t_{1,2}}/M $ (left) and $m_{\tilde t_{3,4}}/M $  (right) plotted in the $m/M$ -- $\tilde m_{F}^2/M^2$ plane, assuming $\tilde m_{f}^2 = 0$.  
}
\label{fig:Pstop}
\end{figure*}

Assuming diagonal soft SUSY breaking masses, the corresponding $6\times 6$ scalar mass-squared matrix, in the basis $(q, Q, U, \bar u^*, \bar Q^*, \bar U^*)$, is given by
\begin{eqnarray}
M_S^2  &= & 
 {\rm diag} \left(M_F M_F^\dagger, M_F^\dagger M_F \right) \\
&&+\;
 {\rm diag} \left(\tilde m_q^2, \tilde m_Q^2, \tilde m_U^2, \tilde m_{\bar u}^2, \tilde m_{\bar Q}^2, \tilde m_{\bar U}^2 \right),
\label{eq:MS2}
\end{eqnarray}
where $\tilde m$s are soft SUSY breaking scalar masses of corresponding fields. 
We neglect soft SUSY breaking trilinear couplings,  $b$ terms, the $\mu$ term and electroweak $D$ terms which are all assumed to be of order the EW scale.\footnote{Vectorlike families  were previously considered in connection with naturalness of EWSB because additional large Yukawa couplings  increase the Higgs boson mass~\cite{Babu:2008ge, Martin:2009bg}. However, extra Yukawas also contribute to the running of $\tilde m_{H_u}^2$ and the net benefit is not dramatic~\cite{Martin:2009bg}. We use vectorlike fields to generate  stop masses.}

For simplicity, in what follows we assume: 
$m_q = m_u \equiv m$, $M_Q = M_U \equiv M$, 
$\tilde m_q^2 = \tilde m_{\bar u}^2 \equiv \tilde m_f^2$, $\tilde m_Q^2 = \tilde m_{\bar U}^2 \equiv \tilde m_F^2$ and $\tilde m_U^2 = \tilde m_{\bar Q}^2 \equiv \tilde m_{\bar F}^2$.
These assumptions are not crucial for our discussion.

\subsection{Top quark mass and fermion spectrum}
Masses of three Dirac fermions, that can be obtained by rotating matrix (\ref{eq:MF}) into mass eigenstate basis, are approximately given by: 
$\lambda v_u M^2/(m^2 + M^2)$, $(M^2+m^2)^{1/2}$,   $(M^2+m^2)^{1/2}$, where the corrections to the smallest mass are  $ {\cal O}(\lambda^3 v_u^3/M^2)$ and the two heavy eigenvalues are split by $ {\cal O}(\lambda v_u)$, assuming that $m$ and  $M$ are of the same order.
In the limit of no mixing, $m\to 0$, we recover the expected result, $m_{\rm top} = \lambda v_u$ and two heavy fermions have masses $M$.  For nonzero $m$ and a fixed Yukawa coupling, the measured value of the top quark mass imposes a relation between $m$ and $M$.

The top Yukawa coupling is given by $y_t = \lambda  M^2/(m^2 + M^2)$, the flavor diagonal couplings to heavy quarks are $\pm \lambda  m^2/(2m^2 +2 M^2)$ and the flavor violating couplings between heavy quarks and the top quark are generated (detailed discussion, although in the lepton sector and in different basis, can be found in  Refs.~\cite{Dermisek:2013gta, Dermisek:2015oja}). The  ratio of $m/M$ required to reproduce the top quark Yukawa coupling at the scale where heavy quarks are integrated out, $Q= (M^2+m^2)^{1/2}$, for $\lambda =1\pm0.1$  is plotted in Fig.~\ref{fig:RG} together with the RG evolution of the top Yukawa.

\subsection{Spectrum of scalars}
Assuming equal vectorlike masses  and soft masses of doublets and singlets
highly simplifies the discussion of the spectrum of scalars because the mass eigenvalues become doubly degenerate. Furthermore, neglecting the contribution from  Yukawa coupling, the masses  squared of scalars are:
\begin{flalign}
& m_{\tilde t_{1,2}}^2 = \frac{1}{2} \tilde M^2 - \frac{1}{2} \sqrt{\tilde M^4 - 4  (M^2 \tilde m_f^2 + m^2 \tilde m_F^2 +  \tilde m_f^2 \tilde m_F^2) },\nonumber \\
& m_{\tilde t_{3,4}}^2 = \frac{1}{2} \tilde M^2 + \frac{1}{2} \sqrt{\tilde M^4 - 4  (M^2 \tilde m_f^2 + m^2 \tilde m_F^2 +  \tilde m_f^2 \tilde m_F^2) },\nonumber  \\
& m_{\tilde t_{5,6}}^2  = M^2+m^2 + \tilde m_{\bar F}^2,
\end{flalign}
  where $\tilde M^2 \equiv M^2+m^2 + \tilde m_{f}^2+  \tilde m_{F}^2$.   
The crucial observation is that all scalars acquire masses even if  $\tilde m_{f}^2 = 0$. The $m_{\tilde t_{1,2}} $ and $m_{\tilde t_{3,4}} $ normalized to $M$ are plotted in the $m/M$ -- $\tilde m_{F}/M$ plane, assuming $\tilde m_{f}^2 = 0$, in Fig.~\ref{fig:Pstop}.

\section{ One-loop RG evolution and threshold corrections}
Let us neglect contributions from gaugino masses and $A$ terms and assume that soft masses squared of scalars that couple to $H_u$ are small at the mediation scale,  for  simplicity $\tilde m_{H_u}^2 = \tilde m_f^2 = 0$. Then  in the RG evolution, at one-loop order, $m_{H_u}^2$ and $ \tilde m_f^2$ will remain zero for arbitrarily large soft masses of the  other fields, $\tilde m_F$ and $ \tilde m_{\bar F}$, since these do not couple to $H_u$.
Sufficiently large $\tilde m_f^2$ can be generated by mixing with vectorlike quarks as discussed above without contributing to $m_{H_u}^2$ over a large range in the energy scale. This completely eliminates the largest source of fine-tuning in the EWSB.

\begin{figure*}[th]
\includegraphics[width=2.6in]{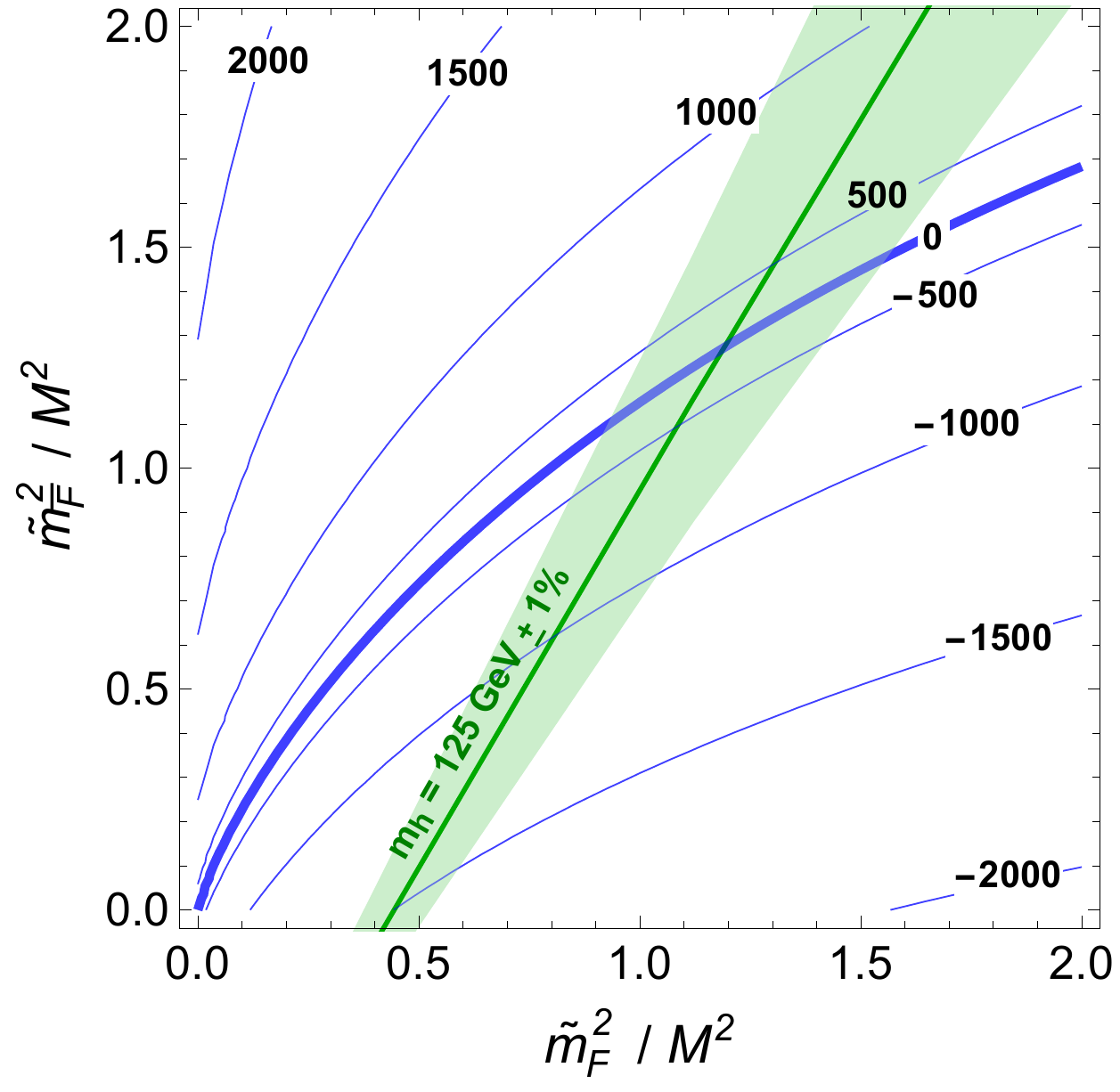}
\hspace{0.5cm}
\includegraphics[width=2.6in]{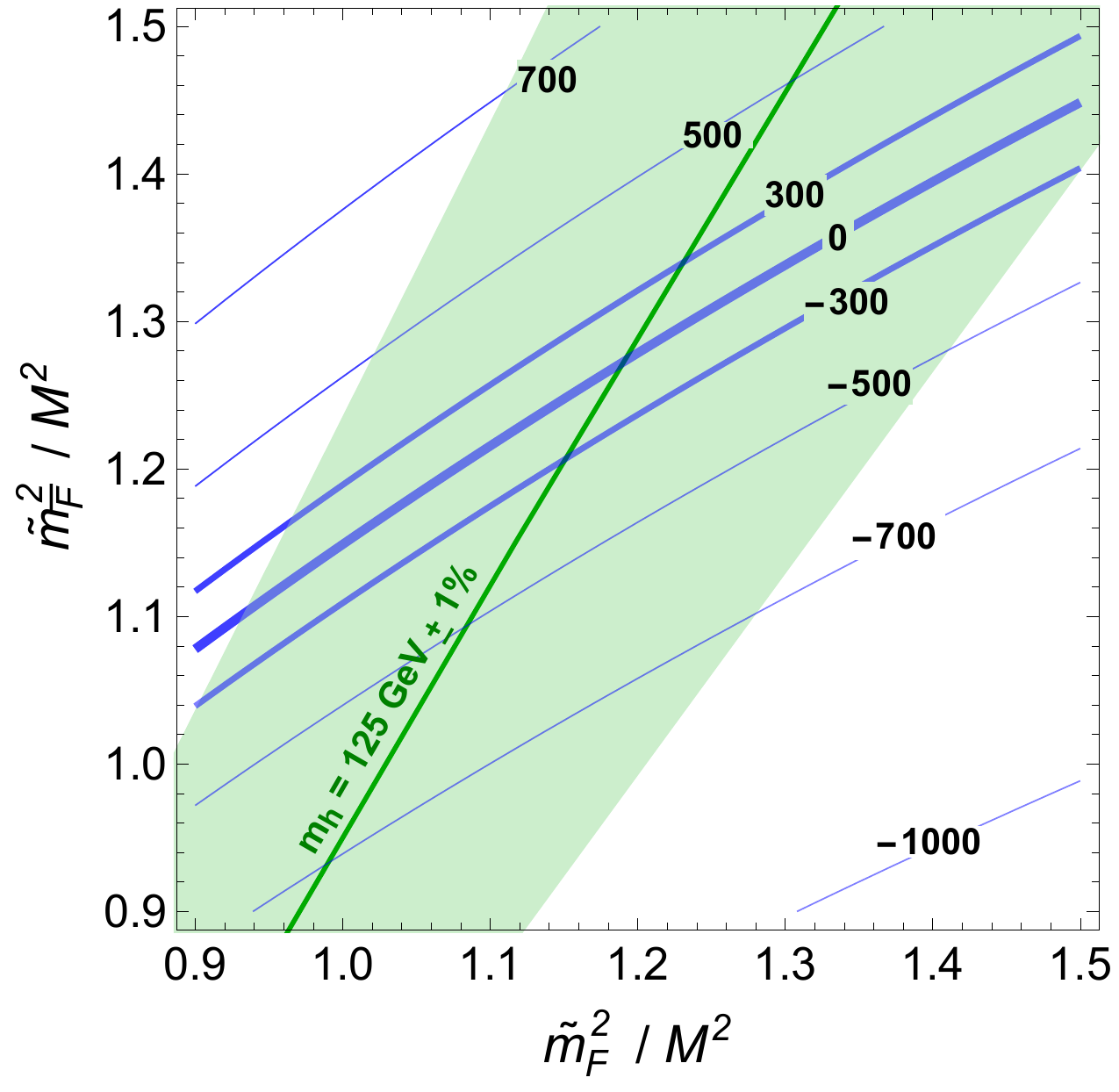}
\caption{Contours of constant contribution to $\tilde m_{H_u}^2/|\tilde m_{H_u}^2|^{1/2}$ [GeV]  from threshold corrections plotted  in the  $\tilde m_F^2/M^2$ -- $ \tilde m_{\bar F}^2/M^2$ plane, for $M = 23$ TeV, $\lambda = 1$ ($m$ is fixed by the top quark mass) and $\tilde m_{f}^2 = 0$.
Along the green line (and shaded area) $m_h = 125$ GeV ($\pm 1\%$)  in our approximation. The matching scale is $Q = m_{\tilde t_{1,2}}$ ($\simeq 9$ TeV in this case).
}
\label{fig:EWSB}
\end{figure*}

Near the $(M^2 +m^2)^{1/2}$ scale, the heavy fermions and all scalars are integrated out. Because of the mixing that generates masses for $\tilde t_{1,2}$, heavy mass eigenstates (both fermions and scalars) acquire couplings to the $H_u$ and generate threshold corrections to $m_{H_u}^2$. For fixed $M$ and $m$, these  corrections do not depend on the renormalization scale at which heavy particles are integrated out (besides the dependence through Yukawa coupling $\lambda$). The threshold corrections are plotted in  Fig.~\ref{fig:EWSB} in the $\tilde m_F^2/M^2$ -- $ \tilde m_{\bar F}^2/M^2$ plane for $M = 23$ TeV, $\lambda = 1$.

For fixed $\lambda$ and vectorlike masses, the  $\tilde m_F^2$ and $ \tilde m_{\bar F}^2$ are the only free parameters that determine masses of superpartners and thus the mass of the Higgs boson. 
The measured value of the Higgs mass, $m_h = 125$ GeV, is obtained along the green line  and the shaded area represents  $\pm 1\%$ range from the central value. We assume that electroweak gauginos and Higgsino are near the EW scale and we match the SM Higgs quartic coupling evolved according to coupled RG equations including contributions from inos to the Higgs quartic coupling predicted from the full model at the scale $Q = m_{\tilde t_{1,2}}$. At this scale, the prediction includes the SUSY tree level result and threshold corrections from integrating out extra fermions and all scalars. The choice $Q = m_{\tilde t_{1,2}}$ is motivated by threshold corrections being small near this scale, typically $\simeq -0.01$.

From Fig.~\ref{fig:EWSB} we see that threshold corrections to $\tilde m_{H_u}^2$ are typically of order $(1 {\rm \; TeV})^2$ for $\tilde m_F^2,\; \tilde m_{\bar F}^2 \leq (30 {\rm \; TeV})^2$. This is expected since the resulting stop masses are  ${\cal O} (10 {\rm \; TeV})$ and the threshold corrections come with the factor $3y_t^2/(8 \pi^2)$ leading to about an order of magnitude suppression. Thus this scenario, without any further assumptions, typically requires about 1\% tuning in EWSB. 

However it is noteworthy that threshold corrections do not necessarily favor EWSB. They can be both positive or negative and there is a range of parameters where the generated corrections are small. The existence of a region leading to  small corrections to $m_{H_u}^2$ does not automatically mean that there is no tuning associated with this region. However the  assessment  of fine-tuning highly depends on further assumptions about the origin of soft scalar masses, namely whether different soft scalar masses are related or independent parameters.

\begin{figure}[t]
\includegraphics[width=2.6in]{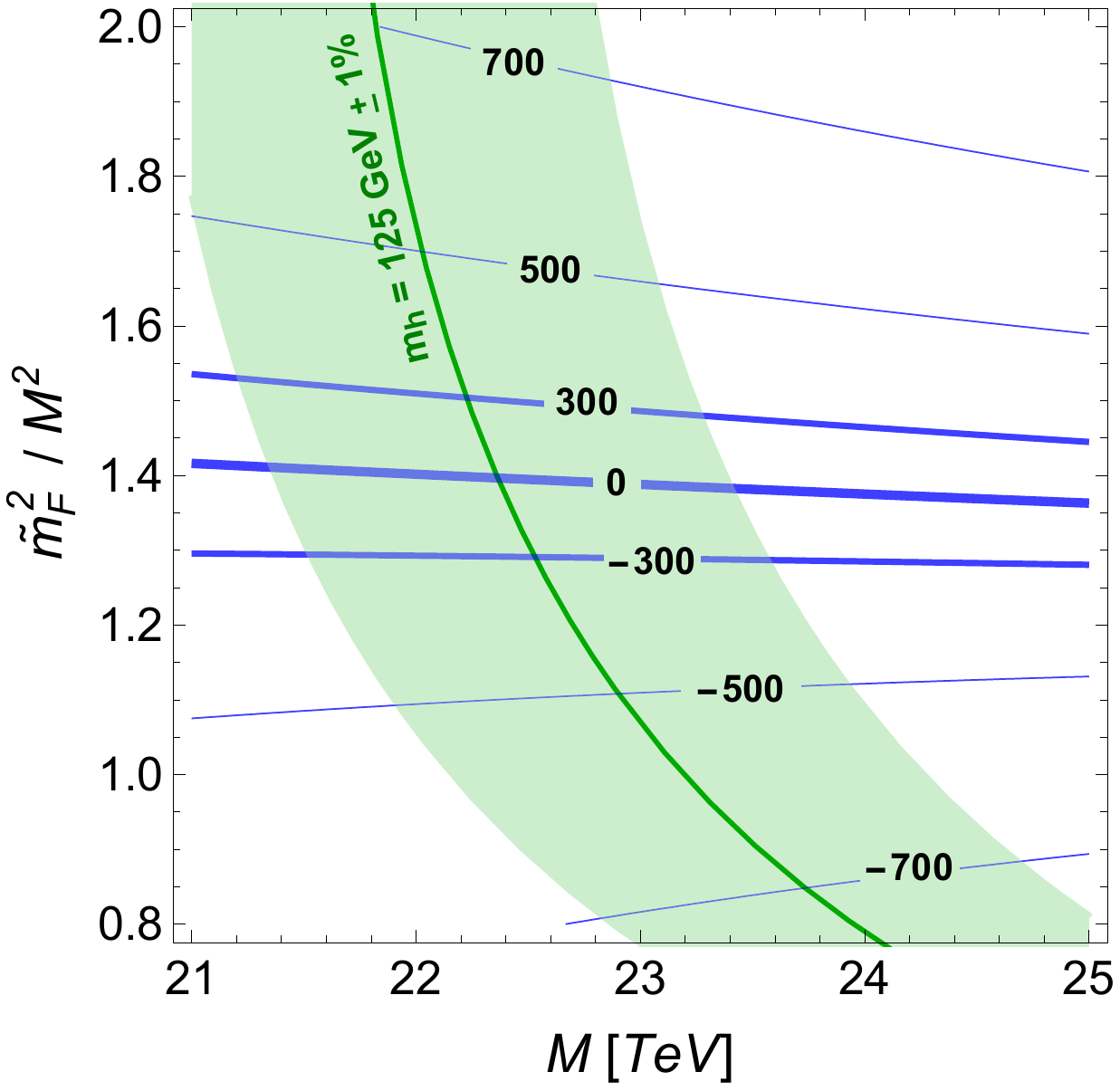}
\caption{The same as in Fig.~\ref{fig:EWSB} plotted   in the  $\tilde m_F^2/M^2$ -- $M$ plane assuming $\tilde m_F^2 =  \tilde m_{\bar F}^2$.
}
\label{fig:EWSBM}
\end{figure}

For example,  each  $\tilde m_F^2$ and  $\tilde m_{\bar F}^2$ represents two soft scalar masses that could be independent parameters. If allowed to vary independently, the contours of   $m_{H_u}^2$ in similar plots to Fig.~\ref{fig:EWSB} would spread by a factor of $\sim\sqrt{2}$. More interestingly, if all soft masses are the same, $\tilde m_F^2  = \tilde m_{\bar F}^2$, the region of parameters with small  $m_{H_u}^2$ is significantly enlarged. This can be understood from Fig.~\ref{fig:EWSB} where $\tilde m_F^2  = \tilde m_{\bar F}^2$ condition implies moving along the diagonal which is almost parallel to contours of $m_{H_u}^2$ in the region of interest. It can also be  seen in Fig.~\ref{fig:EWSBM} where we plot the correction to $m_{H_u}^2$ in the $\tilde m_F^2/M^2$ -- $M$ assuming $\tilde m_F^2  = \tilde m_{\bar F}^2$.

Finally, if soft scalar masses and  vectorlike masses are all related (have a common origin), the contribution to $m_{H_u}^2$ from threshold corrections is controlled by one  mass parameter and small correction to $m_{H_u}^2$ might not require essentially any tuning with respect to that parameter. This is demonstrated in Fig.~\ref{fig:EWSBM}  where contours of constant $m_{H_u}^2$ are almost horizontal lines.

\section{Discussion: two-loop effects and singlet alternatives}

There are two-loop contributions to soft masses of MSSM scalars in the RG evolution originating from heavy scalar masses that can potentially destabilize the hierarchy $\tilde m_f^2 << \tilde m_{F,\bar F}^2$ and significantly affect previous results.  The general form of these two-loop terms is $g^4 {\rm Tr}[\tilde{m}^2]$, where $g$ is a gauge coupling and the trace goes over soft masses squared of all scalars charged under given gauge symmetry~\cite{Martin:1993zk}. The traces of masses squared of $SU(2)$ and $U(1)_Y$ charged scalars affect the $m^2_{H_u}$ directly at the two-loop level, while the trace of masses squared of $SU(3)$ charged scalars contributes to $m^2_{H_u}$ indirectly through contributing to stop masses squared at the two-loop level that, in turn, contribute to $m^2_{H_u}$ at the one-loop level. It turns out that the latter contribution is the dominant two-loop effect for the scenario we discussed. However it is a resummed effect, similar to the contribution from the gluino, and as such it requires evolution over a larger energy interval in order to be effective. 

For the particle content of our scenario, assuming universal heavy scalar masses, the dominant two-loop contribution to stop masses squared from heavy scalars is $-32 (\alpha_3/(4\pi))^2 \tilde m_F^2  \log[\Lambda/\tilde m_F]$, where $\Lambda$ is the mediation scale. It has an opposite sign to the one-loop gluino contribution and these two contributions have equal size for $M_3 = (3\alpha_3/(4\pi))^{1/2} \tilde m_F $. Numerically, 25 TeV heavy scalars contribute approximately as much as a 4 TeV gluino would. 
In order for this contribution not to generate more than $\sim (400 \; {\rm GeV})^2$ correction to $m^2_{H_u}$ and thus not requiring more than $\sim10\%$ tuning,  the mediation scale should not exceed $\sim 250$ TeV.\footnote{The contribution to stop masses squared from heavy scalars for this mediation scale is $-(2 \; \rm{TeV})^2$. Since 10 TeV stop masses in our scenario originate mostly from the mixing with $\sim 25$ TeV scalars, this is a small correction.  Furthermore, this contribution to stop masses is partially canceled by the contribution from gluino.} 
In comparison, the 10 TeV stops in the MSSM, assuming the same mediation scale, would generate  $\sim (3 \; {\rm TeV})^2$ contribution to $m^2_{H_u}$ requiring $\sim0.1\%$ tuning in EWSB. However, as the mediations scale increases, the relative improvement of the scenario with heavy vectorlike quarks compared to the MSSM with 10 TeV stops diminishes.

It should be noted that the two-loop contributions from heavy scalars can be absent if their soft masses squared  come in traceless combinations under every gauge symmetry. Negative soft scalar masses squared for vectorlike fields are not problematic since, due to supersymmetric masses, they do not necessarily lead to tachyons. Not changing any aspect of the scenario we discussed, the easiest possibility would be to introduce additional vectorlike fields that do not couple to the Higgs boson or mix with MSSM fields that have appropriate negative soft masses squared. 

Let us also comment on the scenario where explicit mass terms of vectorlike fields originate from vevs of SM singlets: $m = \lambda_f \langle S_m \rangle$ and $M = \lambda_F \langle S_M \rangle$. Large soft scalar masses squared of heavy fields will drive  the soft scalar masses squared of $S_m$ and $S_M$ in the RG evolution to negative values  
in analogy to the RG evolution of $\tilde m_{H_u}^2$ in the MSSM, see Eq.~(\ref{eq:rge}).  The vevs squared of singlets are related to negative of their masses squared and thus $M^2 \sim m^2 \sim \tilde m_{F,\bar F}^2$ can be achieved. The exact relations will depend on Yukawa couplings $\lambda_{f,F}$ and  couplings from the part of a model that determines quartic couplings of the singlets, which are to a large extent adjustable. However, special attention has to be paid to the $\lambda_f$ coupling because it also generates $\tilde m_{f}^2$ in the RG evolution. In order to preserve the hierarchy $\tilde m_f^2 << \tilde m_{F,\bar F}^2$ in the RG evolution the $\lambda_f$ or  the mediation scale  should not be too large. In addition to  $\lambda_f$, couplings of $S_m$ to other fields in a complete model would also contribute to the RG evolution of its soft mass squared and could make it sufficiently large and negative.

Finally, let us briefly mention an intriguing possibility that the soft masses of heavy fields are generated proportional to their U(1) charges as in D-term mediation of SUSY breaking. Assuming $Q_F = +1$,  $Q_{\bar F} = +1$, $Q_{S_m} = -1$, $Q_{S_M} = -2$ with MSSM fields not charged, the negative soft masses squared of singlets with appropriate sizes are generated directly and in the RG evolution they are not modified due to  $\lambda_{f,F}$ couplings. Similarly, the $\tilde m_{f}^2$ would not be generated in the RG evolution due to  $\lambda_{f}$. Additional vectorlike fields can be added with proper charges to eliminate  two-loop contributions from heavy scalars. Pursuing specific models with a singlet origin of vectorlike masses is beyond the scope of this paper.



\section{Conclusions}

We have discussed a scenario in which   ${\cal O} (10 {\rm \; TeV})$ stops originate from  mixing of states that have a large Yukawa coupling and negligible soft masses and states with no Yukawa coupling but sizable soft masses. As such, the contribution to $\tilde m_{H_u}^2$ generated by large Yukawa coupling to scalars in the RG evolution  from a high scale can be eliminated. The  contribution from scalars is reduced to  threshold corrections and two-loop effects. 

Avoiding a large contribution to $\tilde m_{H_u}^2$ from gluino favors models with low-scale mediation of SUSY breaking. Assuming no specific scenario for generating heavy scalar masses, the two-loop effects from heavy scalars also favor a low mediation scale. However, even for a low  scale the scenario highly reduces the contribution to $\tilde m_{H_u}^2$ from scalar masses. Possibilities to further reduce the two-loop contributions from scalars or remove them completely were outlined.  

It is noteworthy that the EW scale resulting from threshold corrections, with several comparable contributions of both signs, is a prime example of the scenario where the result, significantly smaller than individual contributions, can be understood from the complexity of the model~\cite{Dermisek:2016zvl}.

The mechanism we have discussed does not require any specific relations between parameters and, thus, it can be attached to many models for SUSY breaking.  It can also  be connected with a variety of models that increase the Higgs mass with appropriately lowered scale of vectorlike fields.


\vspace{0.5cm}
\noindent
{\bf Acknowledgments:} R.D. thanks K.S. Babu, H.D. Kim, S. Raby, D. Shih and F. Staub   for useful discussions. This work was supported in part by the U.S. Department of Energy under grant number {DE}-SC0010120
and by  the Ministry of Science, ICT and Planning (MSIP), South Korea, through the Brain Pool Program.



\end{document}